\documentclass[%
twocolumn, aps, 
prb,
]{revtex4-2}

\usepackage[
text={7in,10in},centering,
]{geometry}

\usepackage{amsmath, amsthm, amssymb, calrsfs, wasysym, verbatim, bbm, color, graphics, geometry}
\usepackage{graphicx}
\usepackage{dcolumn}
\usepackage{bm}
\usepackage{hyperref}
\usepackage{amsmath} 
\usepackage{mathtools}
\usepackage{tikz}
\usepackage{tikz-cd}
\usepackage{blindtext}
\usepackage{cleveref}
\usepackage{braket}
\usepackage{siunitx}
\usepackage{lipsum}

\usepackage{floatrow}
\usepackage{graphicx}
\usepackage[label font=bf]{subfig}
\usepackage{caption}
\usepackage[export]{adjustbox}
\graphicspath{ {./images/} }

\usetikzlibrary{quantikz}

\makeatletter
\newcommand*{\rom}[1]{\expandafter\@slowromancap\romannumeral #1@}
\makeatother

\usepackage{changes}
\definechangesauthor[name={WQ Chen}, color=red]{wqc}

\begin{document}

\preprint{APS/123-QED}

\title{Theoretical study of superconducting diode effect in planar \(T_{d}-MoTe_{2}\) Josephson junctions  }

\author{Gongqi Wang}
\affiliation{Department of Physics and Shenzhen Key Laboratory of Advanced Quantum Functional Materials and Devices, Southern University of Science and Technology, Shenzhen 518055, China}

\author{Jianjian Miao}

\author{Wei-Qiang Chen}
\email{chenwq@sustech.edu.cn}
\affiliation{Department of Physics and Shenzhen Key Laboratory of Advanced Quantum Functional Materials and Devices, Southern University of Science and Technology, Shenzhen 518055, China}

\date{August 27, 2024}

\begin{abstract}
	
We investigate the Josephson diode effect (JDE) within quasi-2D planar systems featuring the $C_{1v}$ spin-orbit coupling (SOC) and Zeeman fields in the normal region. Our analysis is based on experimental observations conducted on $MoTe_{2}$ planar Josephson junctions (JJ) subjected to out-of-plane magnetic fields. We emphasize the pivotal role of symmetry breaking in current directionality for the occurrence of the JDE. Specifically, we observe the emergence of asymmetric Andreev bound states (ABSs) and $0$-$\pi$-like transitions with $\varphi_0$-shifts in the current phase relations (CPRs) in systems with specific symmetry breaking induced by SOC and Zeeman fields, leading to different critical current magnitudes in opposite directions. Additionally, we explore the influence of parameters such as the strength of SOC, Zeeman field magnitude and orientation, conduction channels with different transverse momenta, and junction lengths on the JDE efficiencies. Our results indicate the potential for diverse approaches to modulate efficiencies and provide insights that can aid in the discovery of materials and design of Josephson diodes with significantly enhanced efficiency.

\end{abstract}

\maketitle


\section{\label{sec1}INTRODUCTION}

The superconducting diode effect (SDE) is defined by the non-reciprocal flow of supercurrents in opposite directions, a phenomenon distinct from semiconductor diodes which rely on nonreciprocal resistivity. SDE manifests in both junction-free superconductors \cite{Ando2020, Miyasaka2021, Itahashi2020,Schumann2020, Kawarazaki2022, Narita2022, Masuko2022, Lin2022, du2023superconducting} and Josephson junctions  \cite{Baumgartner2021,baumgartner2022effect,Jeon2022,Golod2022,Wu2022,Bauriedl2022,Pal2022,DiezMerida2023,Turini2022,chen2024edelstein}, closely associated with the breaking of inversion and time-reversal symmetry. Its potential applications are akin to those of semiconductors, potentially leading to the development of non-dissipative electronic circuits. Furthermore, the connection between SDE and unconventional superconductors and topological superconductors holds promise for the construction of fault-tolerant quantum computers.

The concept of the Josephson diode was first introduced by Hu et al. in 2007 \cite{hu2007proposed}, leading to the development of various theories on Josephson diode effects (JDE) for different systems, including conventional Josephson junctions (JJs), graphene-based JJs, and topological superconducting JJs. The nonreciprocal Josephson effect associated with the JDE owing to the charging energy difference was studied in \cite{Misaki2021}. JDE can also arise from the valley coupling effect and valley polarization interaction in graphene-based JJs \cite{wei2022supercurrent,hu2023josephson, alvarado2023intrinsic}. The configuration of JJs and specifically designed circuits are alternative methods for producing JDE as well \cite{steiner2023diode,paolucci2023gate,Souto2022,Gupta2023}. Additionally, the origins of JDE could be combination of spin-orbit coupling (SOC), topological materials without inversion symmetry, external magnetic fields or ferromagnets, and the Zeeman effect \cite{Kopasov2021,PhysRevB.105.094517,  davydova2022universal, zhang2022general, pekerten2022anisotropic, tanaka2022theory,Kokkeler2022,   kochan2023phenomenological, Lu2023,Cheng2023,  Liu2024, Huang2023, Fu2024}. Symmetry breaking plays a crucial role in the occurrence of JDE. Many JJ systems have exhibited JDE, with both inversion and time-reversal symmetry being broken by SOCs and magnetic fields or ferromagnets. For instance, the Rashba SOC and in-plane magnetic fields in the normal parts of JJs were discussed by Reynoso et al. \cite{reynoso2012spin}, Yokoyama et al. \cite{yokoyama2014anomalous} and Costa et al. \cite{Costa2023-qu}. 

The relationship between JDE and the emergence of asymmetric Andreev bound states (ABSs) was elucidated in \cite{yokoyama2014anomalous,Costa2023-qu}. When a junction exhibits both inversion and time-reversal symmetry breaking, it may display asymmetric ABSs. Because the current of JJs is related to the change in free energy with respect to the phase difference \cite{Beenakker1991} and the contributions to the supercurrent correspond to the discrete spectrum (ABSs) in the free energy, the asymmetric properties of ABSs are reflected in the $0$-$\pi$-like transition with $\varphi_0$-shift in current–phase relations (CPRs), resulting in different maximum supercurrents in opposite directions.

Recently, SDE was detected in planar Josephson junctions fabricated with \(T_{d}-MoTe_{2}\) \cite{chen2024edelstein}, which is considered to be a type-\rom{2} Weyl semimetal \cite{Jiang2017-ka}.  Research on type-\rom{2} Weyl semimetals has suggested that the JDE arises from the different Fermi velocities of the surface states on the two sides of the material \cite{Chen2018-qq}.  However, in the experiment described in \cite{chen2024edelstein}, the superconducting electrodes don't touch the edges of \(T_{d}-MoTe_{2}\) flake, suggesting that factors other than surface states may contribute to the observed JDE. In addition, the experiment demonstrated that as the thickness of the \(T_{d}-MoTe_{2}\) increased, the strength of the JDE decreased due to a symmetry group transition of \(T_{d}-MoTe_{2}\) from $C_{1v}$ to $C_{2v}$ \cite{Tiwari2021-ps,2020magnetoelectric}. This underscores the significant role played by the symmetric properties of \(T_{d}-MoTe_{2}\) in inducing JDE.

In this study, we primarily focuses on ballistic two-dimensional multichannel JJs with an SOC term corresponding to the $C_{s}$($C_{1v}$) point group. Our findings show that a structure with $C_{1v}$ SOC can also exhibit JDE, even in the absence of the surface state. As few-layer \(T_{d}-MoTe_{2}\) has $C_{1v}$ symmetry, we can introduce a quasi-2D Hamiltonian with $C_{1v}$ SOC to investigate such systems. We present a numerical investigation by discretizing the continuous Hamiltonian and Fourier transform along the current direction to calculate the ABS and CPR. Our results reveal that SDE occurs in the planar JJs of \(T_{d}-MoTe_{2}\) when an out-of-plane magnetic field is applied perpendicular to the direction of the current. We show the emergence of asymmetric ABSs and $0$-$\pi$-like transitions with $\varphi_0$-shifts in the CPRs when both $C_{1v}$ SOC and Zeeman fields are presented. These phenomena result in different magnitudes of the critical currents in opposite directions. The results in \cite{reynoso2012spin,yokoyama2014anomalous,Costa2023-qu} have shown that the JDE is enhanced by the joint effect of ABSs possessing different transverse momenta, observed in both short and long junction limits. However, our study demonstrates the potential enhancement of the JDE arising from the joint effect of multiple ABSs with identical transverse momenta in JJs characterized by intermediate finite junction lengths. The strength of JDE can be quantified by its efficiency
\begin{equation}
	\label{eq2}
	\eta \equiv \frac{J_{\mathrm{c}}^{+}-\left|J_{\mathrm{c}}^{-}\right|}{J_{\mathrm{c}}^{+}+\left|J_{\mathrm{c}}^{-}\right|}.
\end{equation}
where $J_{\mathrm{c}}^{+}$ and $J_{\mathrm{c}}^{-}$ represent the critical densities of supercurrent in opposite directions. We examine the efficiency of the JDE for single and multiple transverse momenta, various SOC strengths, Zeeman fields, and junction lengths.  We find that $\eta$ is not a monotonic function of the above parameters, which is consistent with the experimental results. This result suggests that a systematic study must be conducted to determine the parameters for the best $\eta$. 

The paper is organized as follows. Sec. \ref{sec2} details our model for a 2D planar Josephson junction and outlines our method for computing ABSs and CPRs. In Sec. \ref{sec3}, we present our numerical findings. Sec. \ref{A} investigates the influence of SOC and Zeeman fields on ABSs and CPRs, elucidating the origins of the JDE. We further illustrate the efficiency results for single and multiple transverse momenta, as well as the effects of varying the SOC strength, Zeeman field magnitude and orientation, and junction lengths in Sec. \ref{B} to Sec. \ref{D}.  Finally, Sec. \ref{sec4} summarizes our findings and draws conclusions.

\section{\label{sec2}Model and Method}

\begin{figure}
	\includegraphics[width = 1.0\linewidth]{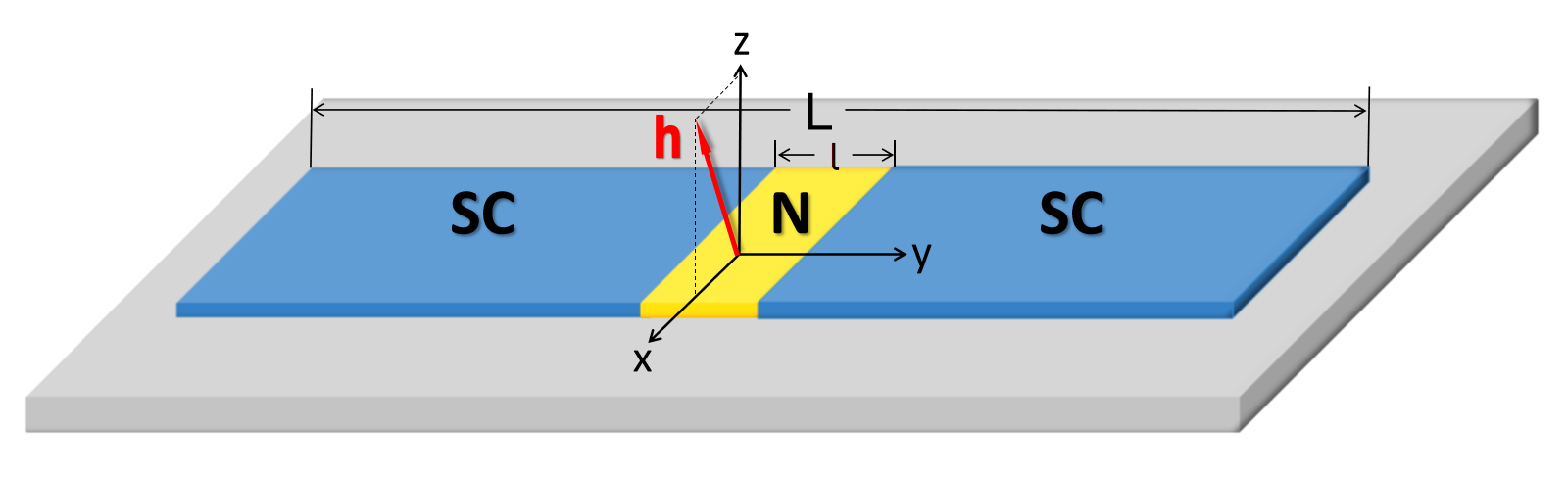}
	\caption{Quasi-2D Josephson junction. The normal part consists of several layers of $T_{d}-MoTe_{2}$ with $C_{1v}$ SOC. We label the junction length by $l$ and the total length with superconducting leads by $L$. The Zeeman field in $xz$-plane is labeled by $\mathbf{h}$. }
	\label{JJs04}
\end{figure}

We examine a 2D Josephson junction illustrated in Fig.\ref{JJs04}. This setup consists of two superconducting electrodes separated by a normal region of a $T_{d}-MoTe_{2}$ flake. The superconducting electrodes are BCS superconductors without SOC or Zeeman fields.

To describe the normal region composed of the quasi-2D planar structure of \(T_{d}-MoTe_{2}\), we begin with a model expressed as
\begin{equation}
	\mathcal{H}=\sum_{\boldsymbol{k}} \psi_{\boldsymbol{k}}^{\dagger} H(\boldsymbol{k}) \psi_{\boldsymbol{k}},
\end{equation}
where $\psi^{\dagger}_{\boldsymbol{k}}=(c^{\dagger}_{\boldsymbol{k} \uparrow}, c^{\dagger}_{\boldsymbol{k}\downarrow})$ is a fermionic spinor and $H(\mathbf{k})$ can be expressed as \cite{du2023superconducting}
\begin{equation}
	\label{eq3}
	\begin{aligned}
	   H(\boldsymbol{k})&=\left[t( k_x^2+ k_y^2)-\mu)\right]\sigma_0+
	   \alpha_1 k_y \sigma_x+\alpha_2 k_x \sigma_y\\
    &\qquad +\alpha_3 k_y \sigma_z +\boldsymbol{h} \cdot \boldsymbol{\sigma}.
    \end{aligned}
\end{equation}
The first term refers to the kinetic energy, where the chemical potential is denoted by $\mu$. The next three terms are the SOC terms with strengths $\alpha_1,\alpha_2$ and $\alpha_3$, and the final term is the Zeeman field term with $\boldsymbol{h}=(h_{x},h_{y},h_{z})$. The Pauli and identity matrices, $\sigma_i (i=x, y, z)$ and $\sigma_0$, act on the spin degrees of freedom. The SOC corresponds to the $C_{s}$ ($C_{1v}$) point group \cite{smidman2017superconductivity} (which is the point group for few-layer $T_{d}-MoTe_{2}$ \cite{2020magnetoelectric}). The SOC term with respect to $k_y$ is $k_y(\alpha_1\sigma_x + \alpha_3\sigma_z)$, which has $y$-inverting symmetry $P_y=\sigma_y$ (where $P_y$ is the $y$ component of the inversion operator $P=P_xP_yP_z$). The external magnetic field breaks this symmetry when at least one of $h_x$ and $h_z$ is not zero \cite{he2022phenomenological}. These symmetry breakings in the $y$-direction suggest the occurrence of SDE or JDE. The Hamiltonian with $h_z\neq 0$ describes the system with an out-of-plane magnetic field in the experiment reported in \cite{chen2024edelstein}.

The current flows along the $y$-direction through the junction. It is assumed that the flake is homogeneous along the $x$-axis. Because of the absence of translational symmetry along the $y$-direction for the entire structure with superconductor electrodes, we place the Hamiltonian on a lattice and consider a tight-binding model. To accomplish this, we substitute $k_{d}^{2}$ with $\frac{2}{a^{2}}(1-\cos(k_{d}a))$ and $k_{d}$ with $\frac{1}{a}\sin(k_{d}a)$ (for $d=x,y$), where $a$ is the lattice constant. For simplicity, we take $a=1$ in the following calculations.  Then we turn $H(\boldsymbol{k})$ into a tight-binding Hamiltonian as
\begin{align}
    \label{eq68}
      H_{tb}(\boldsymbol{k})=&\left[2 t \left(2-\cos k_x - \cos k_y \right)-\mu \right]\sigma_0 + \alpha_1 \sin k_y  \sigma_x \nonumber \\
		&+ \alpha_2 \sin k_x \sigma_y + \alpha_3 \sin k_y  \sigma_z +\boldsymbol{h} \cdot \boldsymbol{\sigma}.  
\end{align}
We take periodic boundary conditions along the $x$-direction, hence $k_x$ is still a good quantum number.  We only need to perform Fourier transform along the $y$-direction, and the Hamiltonian reads
\begin{equation}
	\mathcal{H}=\sum_{ij}\sum_{k_x} \psi_{i,k_x}^{\dagger} h_{i j}(k_x) \psi_{j,k_x},
\end{equation}
where $\psi^{\dagger}_{i,k_x}=(c^{\dagger}_{ik_x \uparrow}, c^{\dagger}_{ik_x\downarrow})$ and the matrix $h_{i j}(k_x)$ with the lattice index $i$ and $j$ in the Hamiltonian is
\begin{align}
	\label{eq73}
		h_{i j}(k_x)= & \{\left[2 t \left(2-\cos k_x\right)-\mu\right] \sigma_0 + \alpha_2 \sin k_x \sigma_y\} \delta_{ij} \nonumber \\
		&-t \sigma_0 \left(\delta_{i+1, j}+\delta_{i-1, j}\right)  \nonumber\\
		&+\frac{1}{2 i}\left(\alpha_1 \sigma_x+ \alpha_3 \sigma_z\right) \left(\delta_{i+1, j}-\delta_{i-1, j}\right) \nonumber \\
		&+\boldsymbol{h} \cdot \boldsymbol{\sigma}\delta_{ij}.
\end{align}

For simplicity, we assume that the kinetic energy of the superconducting electrodes is also given by equation \eqref{eq73} but without the SOC and Zeeman terms.  Then the Bogoliubov-de Gennes (BdG) Hamiltonian of the entire setup is given by 
\begin{equation}
	\label{eq82}
	H_{ij}(k_x)=\left(\begin{array}{cc}
		h_{ij}(k_x) & \Delta\sigma_0 \delta_{ij} \\
		\Delta^{\dagger}\sigma_0 \delta_{ij} & -\hat{T}\left[h_{ij}(-k_x) \right] \hat{T}^{-1}
	\end{array}\right),
\end{equation}
where $\hat{T}=-i \sigma_y \mathcal{K}$ is the time-reversal operator.
The total length of the structure illustrated in Fig.\ref{JJs04} is $L$, and the length of the junction is $l$. The gap functions of the superconducting electrodes on the left and right sides are $\Delta=\Delta_0$ and $\Delta_0 e^{i \varphi}$, respectively, where $\varphi$ represents the relative phase difference.
Here, we focus solely on the s-wave pairing of the superconducting components. The strengths of all three SOC parameters are set to be the same ($\alpha_1=\alpha_2=\alpha_3 \equiv \alpha$) in the normal region for all the calculations. The total Hamiltonian can be expressed as
\begin{equation}
	\label{eq80}
	\mathcal{H}_{BdG}=\sum_k \sum_{i, j} \Psi_{i, k}^{\dagger} H_{ij} \Psi_{j, k},
\end{equation}
where $\Psi_{i, k_x}^{\dagger}=\left[c_{i k_x \uparrow}^{\dagger}, c_{i k_x \downarrow}^{\dagger}, c_{i -k_x \downarrow}, -c_{i -k_x \uparrow}\right]$ is the Nambu spinor. By diagonalizing the BdG equation numerically, one can obtain the spectrums $\{E_n(\varphi)\}$ of quasiparticles. 
The ground state energy of the JJ system is expressed as
\begin{equation}
	\label{eq66}
	\varepsilon(\varphi)=-\frac{1}{2} \sum_n{}  E_n(\varphi).
\end{equation}
The calculation of the Josephson current density is based solely on the sum of the positive ABSs which are the energy levels in the range of $0 \leqslant E_n \leqslant \Delta_0$. The formula for determining the Josephson current density is as follows:
\begin{equation}
	\label{eq65}
	J(\varphi)=\frac{2 e}{\hbar} \frac{\mathrm{d} \varepsilon(\varphi)}{\mathrm{d} \varphi}.
\end{equation}
The maximum supercurrent density in the positive and negative directions, $J_c^{+}$ and $J_c^{-}$ can be obtained from $J(\varphi)$. In a quasi-2D system, quasiparticles can possess diverse transverse momenta $k_x$, which are also referred to as conduction channels or $k_x$-channels. The total supercurrent can be calculated by summing all the supercurrent densities with each allowed $k_x$-channel.


\section{\label{sec3}Results and discussions}

\subsection{\label{A}ABSs and CPRs}
\begin{figure}
		\subfloat{\includegraphics[width= 0.5\linewidth,valign=b]{
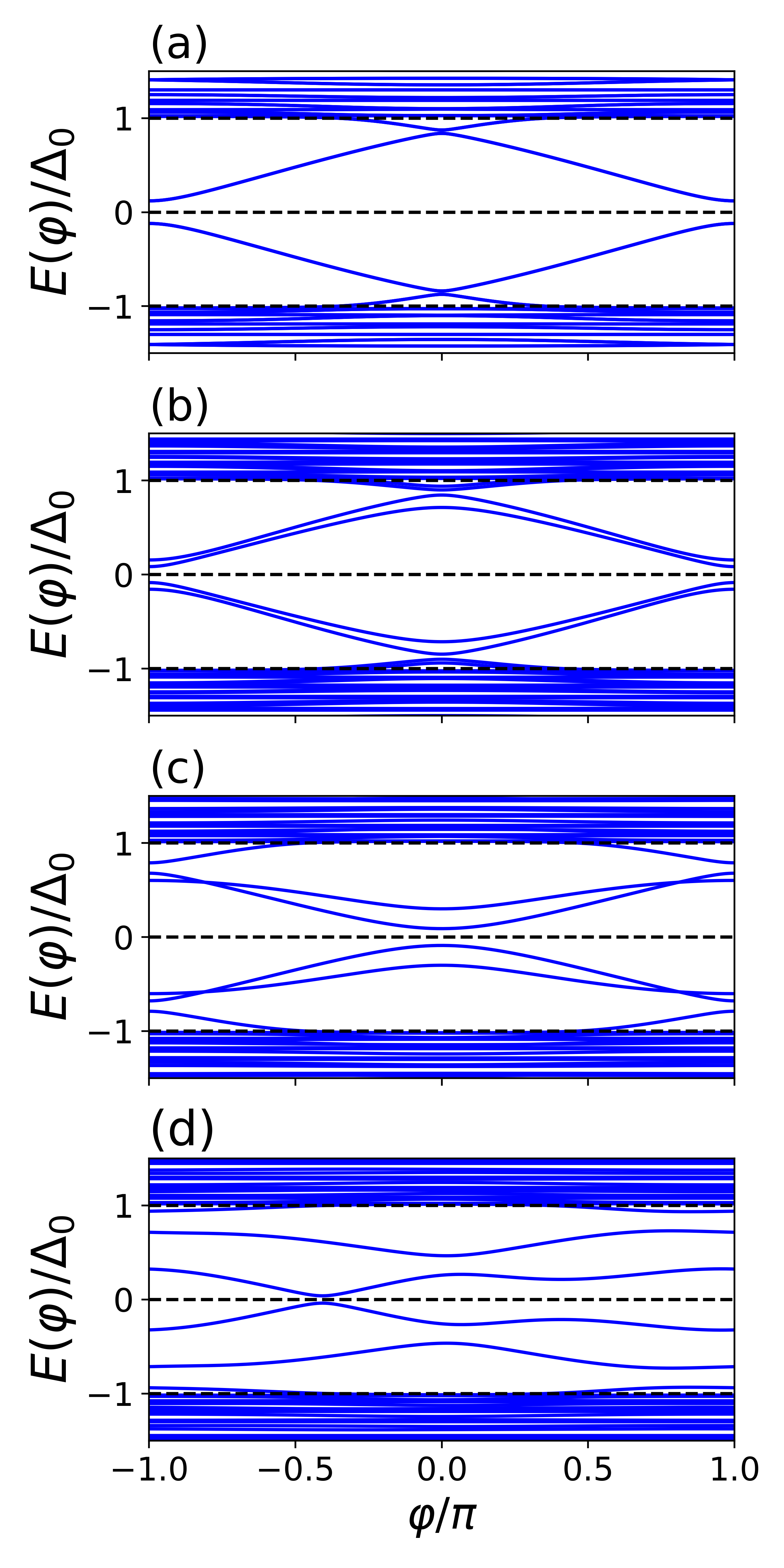}\label{}}
		\subfloat{\includegraphics
[width= 0.5\linewidth,valign=b]{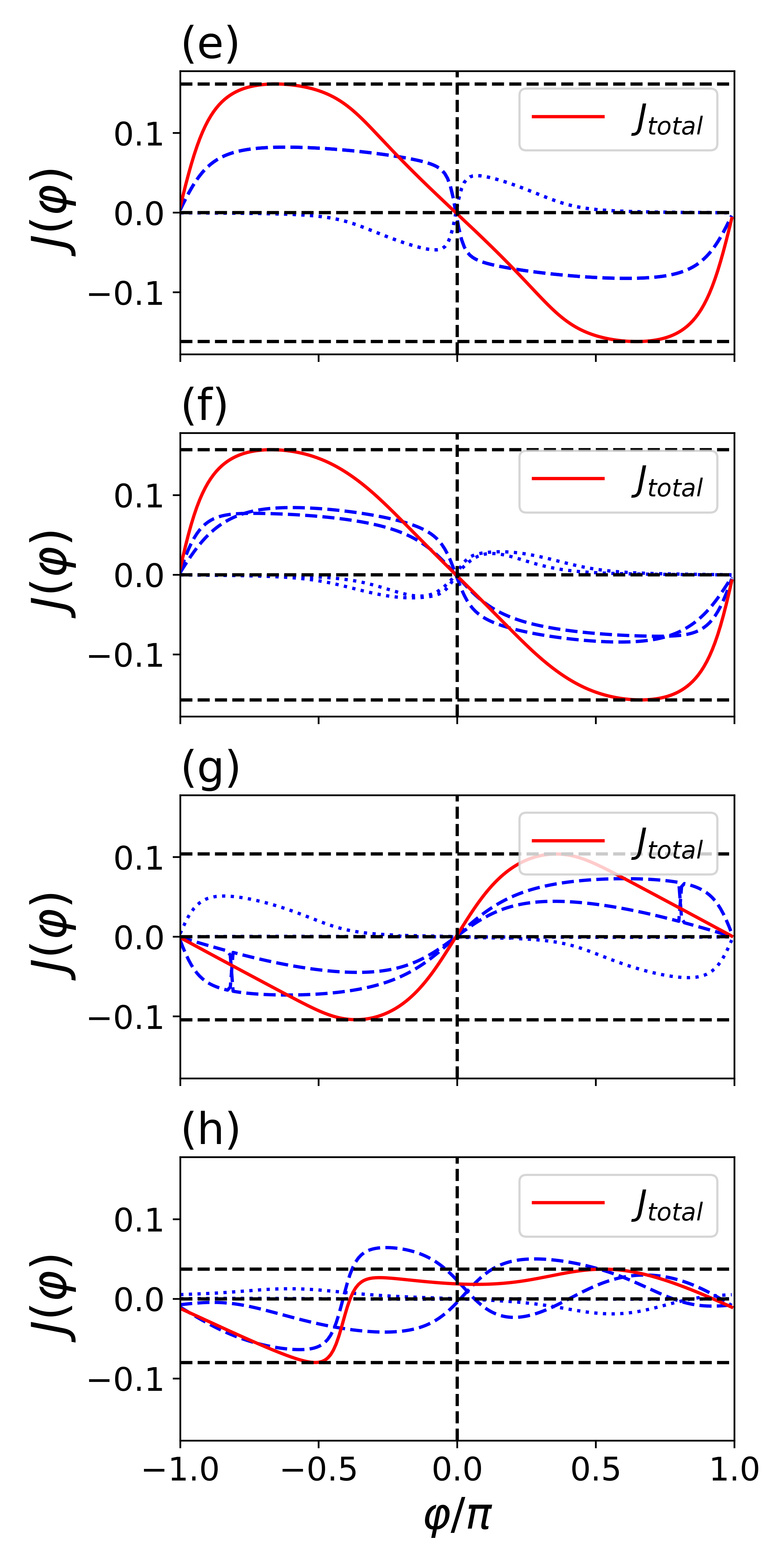}\label{}}\\
	\caption{ABS (in between energy gap) at $k_x=0.5$ with (a) $\alpha=0$ and $h_z=0$, (b) $\alpha=0.6$ and $h_z=0$, (c) $\alpha=0$ and $h_z=0.6$, and (d) $\alpha=0.6$ and $h_z=0.6$. (e), (f), (g), and (h) are the CPRs corresponding to the left-hand ABSs. The blue dashed and dotted lines correspond to the currents from the two lower and upper ABSs, respectively. The red lines represent the total current density from the joint contribution. 
		The other parameters in the calculation are $l=5L/100 $, $L=150$, $a=1$, $t =3$, $\mu=1.5$ and $\Delta_0=0.3$.}
	\label{ABS compare}
\end{figure}

In this section, we explore how the SOC and Zeeman field affect the ABS and supercurrents. For clarity, we consider the contribution of a single channel with a transverse momentum of $k_x=0.5$. We will expand our analysis to include additional $k_x$-channels later. To generate the ABS spectrum in Fig. \ref{ABS compare}, we solve the BdG equation \eqref{eq80} for each fixed value of $\varphi$. The supercurrent depicted in Fig. \ref{ABS compare} is computed using equations \eqref{eq66} and  \eqref{eq65}, taking into account contributions from $0 \leqslant E_n \leqslant \Delta_0$. Fig.\ref{ABS compare} (a) displays the ABS spectrum with double degeneracy in the absence of the SOC term and Zeeman field, along with the corresponding supercurrents in Fig.\ref{ABS compare} (e). The number of ABSs inside the gap increases with the magnitude of the gap and the length of the junction.  For our chosen gap and junction length $l=5L/100$, our calculations involve two doubly degenerate ABSs. The contributions of the upper and lower ABS to the total supercurrent differ, as evidenced by the dashed and dotted blue curves in Fig.\ref{ABS compare} (e)).
 
The SOC term breaks the two-fold spin degeneracy of ABSs at a finite transverse momentum (Fig.\ref{ABS compare} (b)). The magnetic field breaks spin degeneracy at any transverse momentum due to time-reversal symmetry breaking (Fig.\ref{ABS compare} (c)). The presence of a magnetic field can lead to a $0$–$\pi$ transition, where the current sign changes at a finite phase difference $\varphi$, as demonstrated by the transfer of CPRs in Fig.\ref{ABS compare} (e) to (g), consistent with prior research \cite{PismaZhETF25314,PhysRevLett.86.2427,PhysRevLett.96.117005,PhysRevLett.104.117002,KAWABATA20123467,Costa2018-nr}. The CPRs remain odd functions and are anti-symmetric about $\varphi=0$ when the $0$-$\pi$ transition occurs. In the presence of both SOC and the Zeeman field, the ABSs become a complicated function of $\varphi$ and asymmetric about $\varphi=0$ (Fig.\ref{ABS compare} (d)). This asymmetry results in anomalous phase shifts $J(\varphi) \neq J(-\varphi)$ and $J(\varphi=0) \neq 0$ , known as $\varphi_0$-shifts, as depicted in Fig.\ref{ABS compare} (h).

The presence of SOC alters the $0$-$\pi$ transition, resulting in $\varphi_0$-shifts and breaking the anti-symmetry of CPRs, leading to a modified $0$–$\pi$ transition referred to as a $0$–$\pi$-like transition. In the presence of the Zeeman field alone, the positive and negative critical current densities $J_c^{\pm}$ and the corresponding critical phase difference $\varphi_c^{\pm}$ maintain a symmetric property during the $0$-$\pi$ transition. However, in the presence of both the SOC and Zeeman fields, this symmetry is absent, indicating that for a single ABS, $\varphi_c^{\pm}$ can possess the same sign and correspond to different magnitudes of supercurrent density (as indicated by the dashed line in Fig.\ref{ABS compare} (h)). 
The relationship between JDE  and asymmetric ABSs with different $k_x$-channels has been discussed in  \cite{reynoso2012spin} and \cite{yokoyama2014anomalous} for short and long junction limits.
The occurrence of the JDE is attributed to the $0$-$\pi$-like transition with $\varphi_0$-shift caused by the single-split asymmetric ABS in the gap. Our calculations show that the contribution to the total supercurrent density comes from multiple ABSs of a single $k_x$-channel in the gap, rather than from a single ABS as discussed in \cite{Costa2023-qu}. Fig.\ref{ABS compare} (f) and (g) show that the total critical supercurrent densities $|J_c^\pm|=0.157$ and $|J_c^\pm|=0.104$ when only the spin-orbit coupling (SOC) or the Zeeman field is present, suggesting no JDE. However, in the presence of both the SOC and Zeeman fields, as shown in Fig.\ref{ABS compare} (h), we observe the total critical supercurrent densities $|J_c^+|=0.079 \ne |J_c^-|=0.037$, indicating the presence of JDE. 

The joint contribution of multiple ABSs to the total supercurrent density can potentially strengthen the JDE. For example, as depicted in Fig.\ref{ABS compare} (d), the major contribution to the total supercurrent arises from the lower two ABSs, while the ABS near the gap edge makes a minor contribution. This disparity is due to the larger magnitudes of supercurrent density corresponding to the lower two ABSs compared to the upper one across most of the $\varphi$. The respective efficiencies of the three ABSs are $\eta = 0.005, 0.092, -0.193$. The combined magnitude of their efficiencies, $|\eta| = 0.363$ exceeds the individual magnitudes, with the two major contributors being notably smaller. This discrepancy stems from the substantial phase shift differences among the ABSs, resulting in an offset or enhancement of the supercurrent at a given $\varphi$. 
Further discussion on this enhancement is provided in Sec. \ref{D} for varying junction lengths.

Additionally, we observe a shift in the contribution of each ABS to the supercurrents in the presence of spin-orbit coupling (SOC) and the Zeeman field, as evident in the comparison between Fig.\ref{ABS compare} (e) to (f). For example, in Fig.\ref{ABS compare} (g), the uppermost ABS approach and merges into the continuous spectrum, leading to the disappearance of the corresponding supercurrent. This phenomenon is also observed in Fig.\ref{ABS compare} (h). The occurrence of the JDE consistently coincides with changes in contributions, as split ABSs approach or move away from the energy gap's edge. With an increase in the magnetic field, some ABSs may transition into the continuous spectrum, while certain parts of the continuous spectrum may enter the gap. 

In brief, the emergence of a $0$-$\pi$-like transition with a $\varphi_0$-shift from a single-split ABS can engender the JDE, while multiple ABSs from a single $k_x$-channel have the potential to amplify the JDE's strength. Furthermore, the mixed split ABSs from multiple $k_x$-channels can influence the JDE's strength, contingent upon the chosen parameters. These aspects will be further explored in the subsequent subsection, alongside the impact of multiple $k_x$-channels.

\subsection{\label{B}CPRs and efficiencies of different conduction channels}
\begin{figure}
	\subfloat{\includegraphics[width= 1.0\linewidth,valign=b]{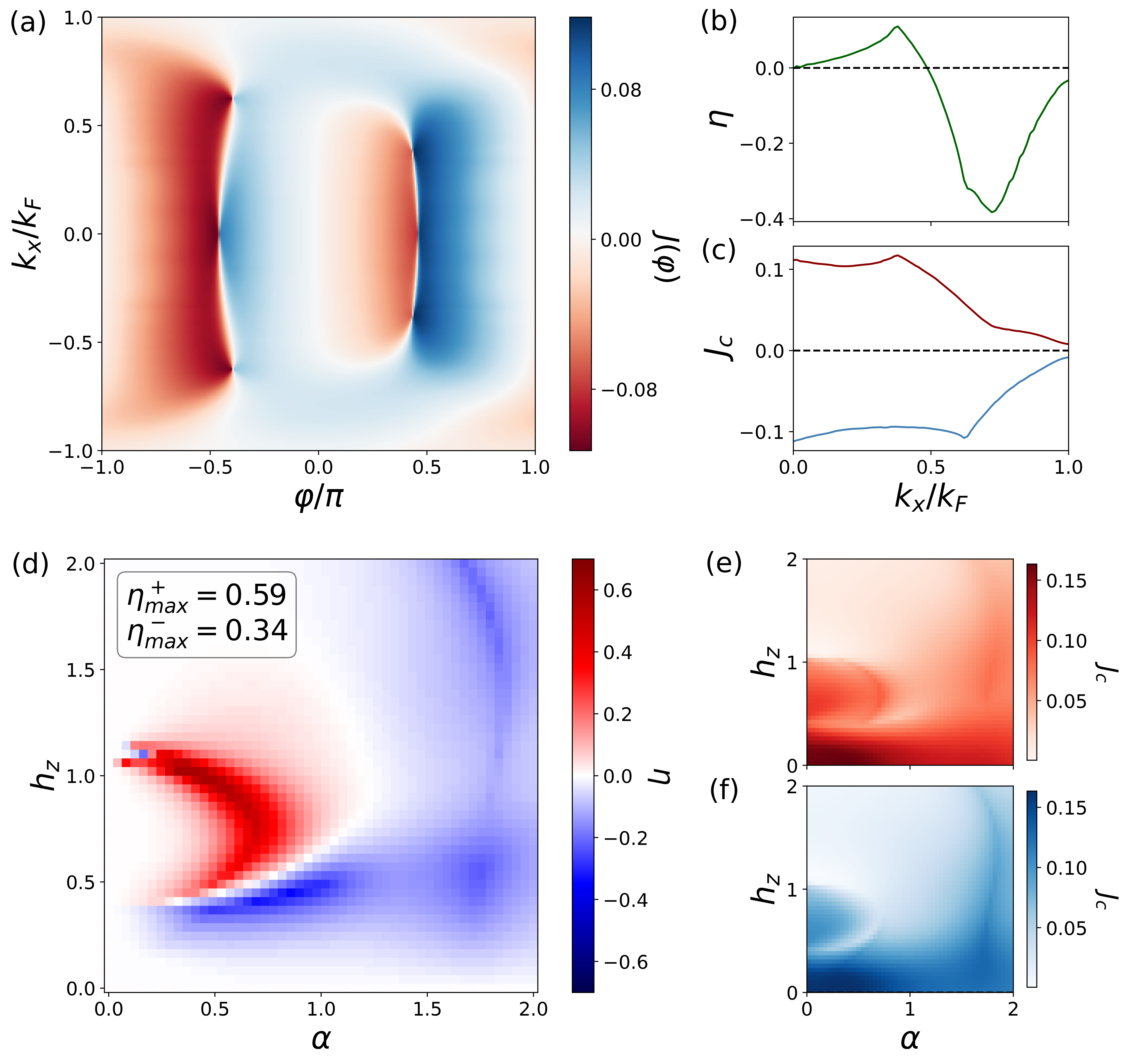}\label{}}\\
\caption{ (a) The color map of CPRs with a dense sampling of $k_x$ and $\varphi$. (b) The efficiencies as a function of $k_x$. (c) The positive (red line) and negative (blue line) critical currents as functions of $k_x$. The SOC strength is $\alpha=0.6$ and Zeeman field is $h_z=0.6$ for (a), (b) and (c). (d) The color map of efficiencies, (e) positive and (f) negative critical currents  with a dense sampling of $\alpha$ and $h_z$. $k_x=0.5$ for (d), (e) and (f). $\eta^+_{max}$ and $\eta^-_{max}$ are the maximum magnitudes of positive and negative efficiencies.
The other parameters in the calculation are $l=5L/100 $, $L=150$, $a=1$, $t =3$, $\mu=1.5$ and $\Delta_0=0.3$.}
\label{multikxeffcon}
\end{figure}

In this subsection, we first discuss the CPR and efficiency of JDE for different channels with the same SOC and Zeeman field. In the calculation for Fig.\ref{multikxeffcon} (a), (b), and (c), the strength of SOC $\alpha=0.6$ and Zeeman field $h_z=0.6$ are the same as in the last subsection.

In the previous subsection, we explored the $0$-$\pi$-like transition with the transverse momentum of $k_x=0.5$. In this subsection, we will demonstrate $0$-$\pi$-like transitions for various $k_x$-channels. In Fig.\ref{multikxeffcon} (a), we observe changed $0$-$\pi$-like transitions with respect to $k_x$. Similar results were discussed analytically for a short Josephson junction with Rashba SOC and delta-like magnetic exchange \cite{Costa2023-qu}. This can be explained by the ``renormalized'' channel-specific Zeeman field (exchange parameters in \cite{Costa2023-qu,Costa2018-nr}). The presence of the transverse momenta effectively changed the Zeeman field, leading to distinct phase shifts for the given $k_x$-channels, as indicated by different situations of $0$-$\pi$-like transitions. $\varphi_0$-shifts are shown in Fig.\ref{multikxeffcon} (a) as the magnitude of $k_x$ increases, becoming obvious when $k_x$ is roughly larger than $0.5k_F$. The results in \cite{Costa2023-qu} show a obvious $\varphi_0$-shifts when the magnitude of the transverse momenta is larger than a critical value.
This is due to the fact that the upper split level of the two-fold spin-degenerate ABS approaches the gap edge and transforms into a flat band once transverse momenta exceed a critical value (which corresponds to a relatively large ``renormalized" channel-specific Zeeman field). Meanwhile, the lower split level that is situated inside the gap experiences the most significant impact of time-reversal symmetry breaking, resulting in sizable $\varphi_0$-shifts. This phenomenon can also be seen in Fig.\ref{ABS compare} (d) and (h), and we previously discussed the alteration in the contribution of each ABS to supercurrents in Sec. \ref{A}.
As a result, a similar difference in the $\varphi_0$-shifts for different $k_x$ can be observed in Fig.\ref{multikxeffcon} (a). However, there is no obvious critical $k_x=k^{crit}_x$ for the occurrence of $\varphi_0$-shifts when $|k_x|>|k_x^{crit}|$ because the junction in the presence of the Zeeman field is finite which is more close to real junction  in our calculation but not a delta-like magnetic link.

The efficiency of the SDE was assessed using equation \eqref{eq2}. As shown in Fig.\ref{multikxeffcon} (b), the efficiency $\eta$ exhibits a non-monotonic change with $k_x$ and reaches its maximum magnitude at an optimal $k_x$. Fig.\ref{multikxeffcon} (c) illustrates the disparity between critical currents with opposite directions, and small peaks and kinks are observed in the critical currents as functions of $k_x$. These phenomena can be attributed to the ``renormalized" channel-specific Zeeman field, which induces phase shifts with varying $k_x$ and modifies the critical current as reported in \cite{Costa2023-qu, Costa2018-nr, Yokoyama2014-pj}. Because various values of $k_x$ represent distinct $0$-$\pi$-like transitions, the corresponding strength of the JDE changes, as shown in Fig.\ref{multikxeffcon} (b) by efficiencies.
Note that only positive $k_x$ values are considered here because the CPRs are symmetric about $k_x=0$ (Fig.\ref{multikxeffcon} (a)).
If we want to make use of the large efficiencies appearing at the single optimal transverse channels, the systems have to be a quasi-1D system, allowing only one or a few transverse channels. However, if we consider a quasi-2D system, the joint effect from different transverse channels must be considered, as we will discuss in the next subsection.

\subsection{\label{C}Efficiencies with various strengths of SOC and Zeeman fields}

To investigate the effects of the SOC and Zeeman fields on critical currents and efficiencies, we conducted calculations across a range of parameters. Specifically, the SOC strength ($\alpha$) and Zeeman field ($h_z$) were varied within the range $\alpha,h_z \in [0,2]$.
Initially, we present the results for a single transverse momentum ($k_x=0.5$) as discussed in Sec. \ref{A}. In Fig.\ref{multikxeffcon} (e) and (f), the critical currents in both the positive and negative directions exhibit non-monotonic behavior, attributed to phase shifts caused by the SOC and Zeeman fields. These phase shifts result in oscillations in critical current, as previously discussed in \cite{Yokoyama2014-pj}. Notably, the magnitudes of the critical current in the opposite directions remain equal, despite the phase shift arising from the SOC and Zeeman fields in the 1D Rashba system. This suggests that the 1D Rashba system with SOC does not exhibit SDE, even in the presence of the Zeeman field perpendicular to the current direction. This is because the Rashba SOC term can be removed by a spin-dependent gauge transformation \cite{li2017spin}. Similar findings have been reported for junction-free systems \cite{legg2022superconducting}. However, in the 2D Rashba system or the 1D and 2D $C_{1v}$ systems, a spin-dependent gauge transformation cannot be applied, leading to the observation of SDE and JDE. Fig.\ref{multikxeffcon} (d) illustrates the varying efficiencies of the single transverse momentum $k_x=0.5$ with respect to both the SOC and Zeeman fields, displaying non-monotonic behavior and reaching maximum magnitudes at specific optimal values of the SOC and Zeeman fields. While maintaining a fixed SOC, the efficiencies display a change in sign with varying Zeeman fields, consistent with experimental observations involving \(T_{d}-MoTe_{2}\) JJs \cite{chen2024edelstein}. These properties of the efficiencies suggest significant differences in the phase shifts in the $0$-$\pi$-like transitions in the CPRs for various combinations of SOC and Zeeman fields, emphasizing the need for a system with tunable SOC and Zeeman fields to achieve maximum magnitude of efficiency.
	
\begin{figure}
	\subfloat{\includegraphics[width= 1.0\linewidth,valign=b]{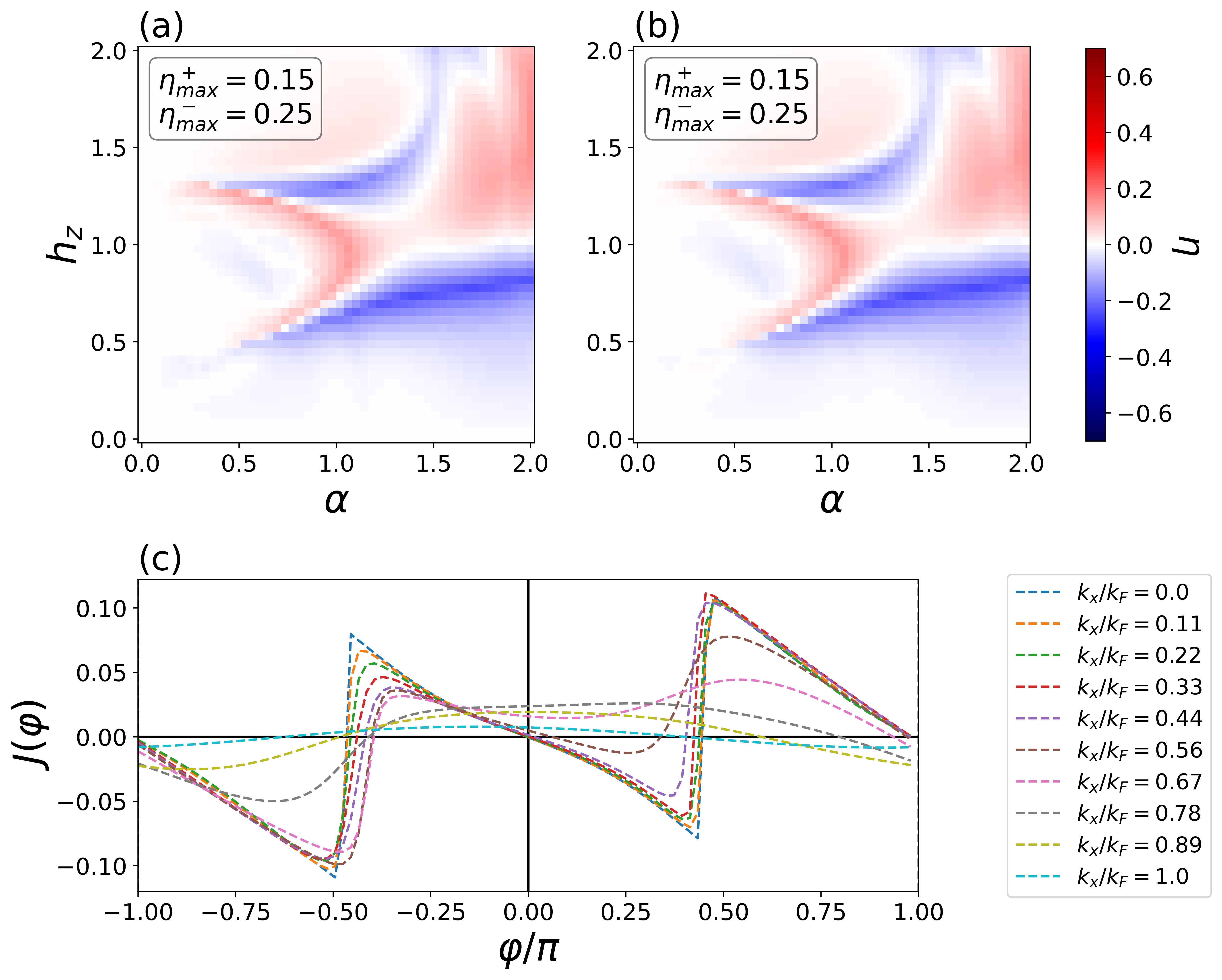}\label{}}\\
	\caption{The color map of efficiencies with (a) $N=10$ and (b) $N=15$ with a dense sampling of $\alpha$ and $h_z$. (c) The CPRs for every $k_x$ in the calculation with $N=10$. The other parameters in the calculation are $l=5L/100 $, $L=150$, $a=1$, $t =3$, $\mu=1.5$ and $\Delta_0=0.3$.}
	\label{multikxeffconcu}
\end{figure}

To explore quasi-two-dimensional systems, it is necessary to determine the sum of the current densities across all the allowed $k_x$-channels. The results are presented by selecting $k_x$ uniformly spaced within the range of $k_x \in [0, k_F]$, with numbers of $k_x$-channels $N=10$ and $N=15$ in Fig.\ref{multikxeffconcu} (a) and (b), respectively. The distributions of the efficiencies differ from those in the single $k_x$-channel case (Fig.\ref{multikxeffcon} (d)). This difference arises because for various $k_x$-channels, the positive and negative critical currents correspond to distinct values of $\varphi$, and the $0$-$\pi$-like transitions are dissimilar. The sum of the current densities for different $k_x$-channels leads to an offset or enhancement at different values of $\varphi$, as  illustrated in Fig.\ref{multikxeffconcu} (g) for $N=10$ at $\alpha=0.6$ and $h_z=0.6$.
For a fixed SOC and Zeeman field, the efficiency may change non-monotonically with an increase in $N$. The patterns shown in Fig.\ref{multikxeffconcu} (a) and (b) indicate that the change in efficiency diminishes as $N$ increases, as the results tend toward the outcome by integrating over all possible $k_x$-channels. Comparing the maximum magnitudes of the efficiencies in Fig.\ref{multikxeffconcu} (a) and (b) with Fig.\ref{multikxeffcon} (d) reveals that a single $k_x$-channel can correspond to a larger efficiency at the optimal value of the SOC and Zeeman fields, with maximum magnitudes of $\eta_{max} = 0.25, 0.25, 0.59$, respectively.
Furthermore, it can be inferred that for the same material with a fixed SOC, Zeeman field, and Fermi level, the efficiency may change non-monotonically with an increase in the junction width and approach a constant value, as wider junctions allow more $k_x$-channels.

\subsection{\label{D}Efficiencies with different Zeeman field magnitudes and orientations; different junction lengths}

\begin{figure}[h]\centering
	\subfloat{\includegraphics[width= 1.0\linewidth,valign=b]{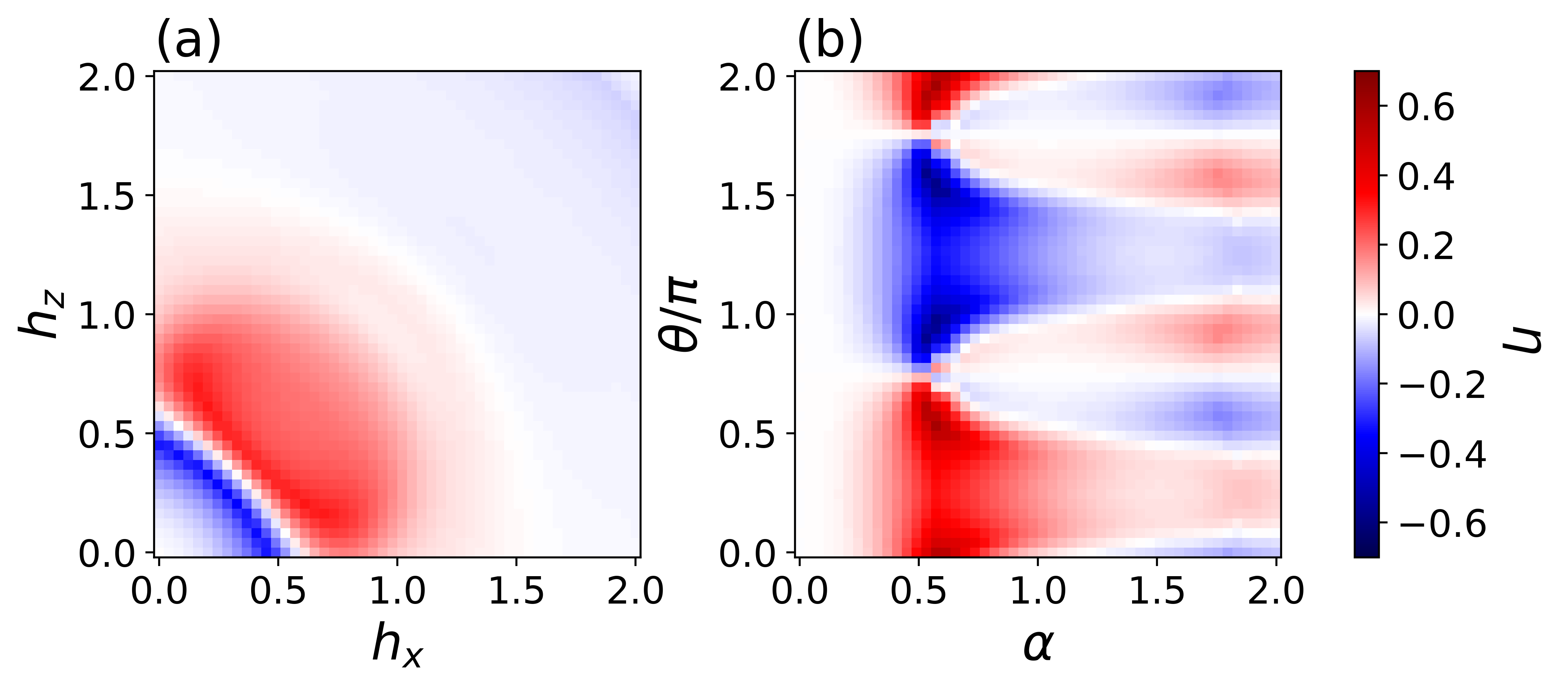}\label{}}\\
	\subfloat{\includegraphics[width= 1.0\linewidth,valign=b]{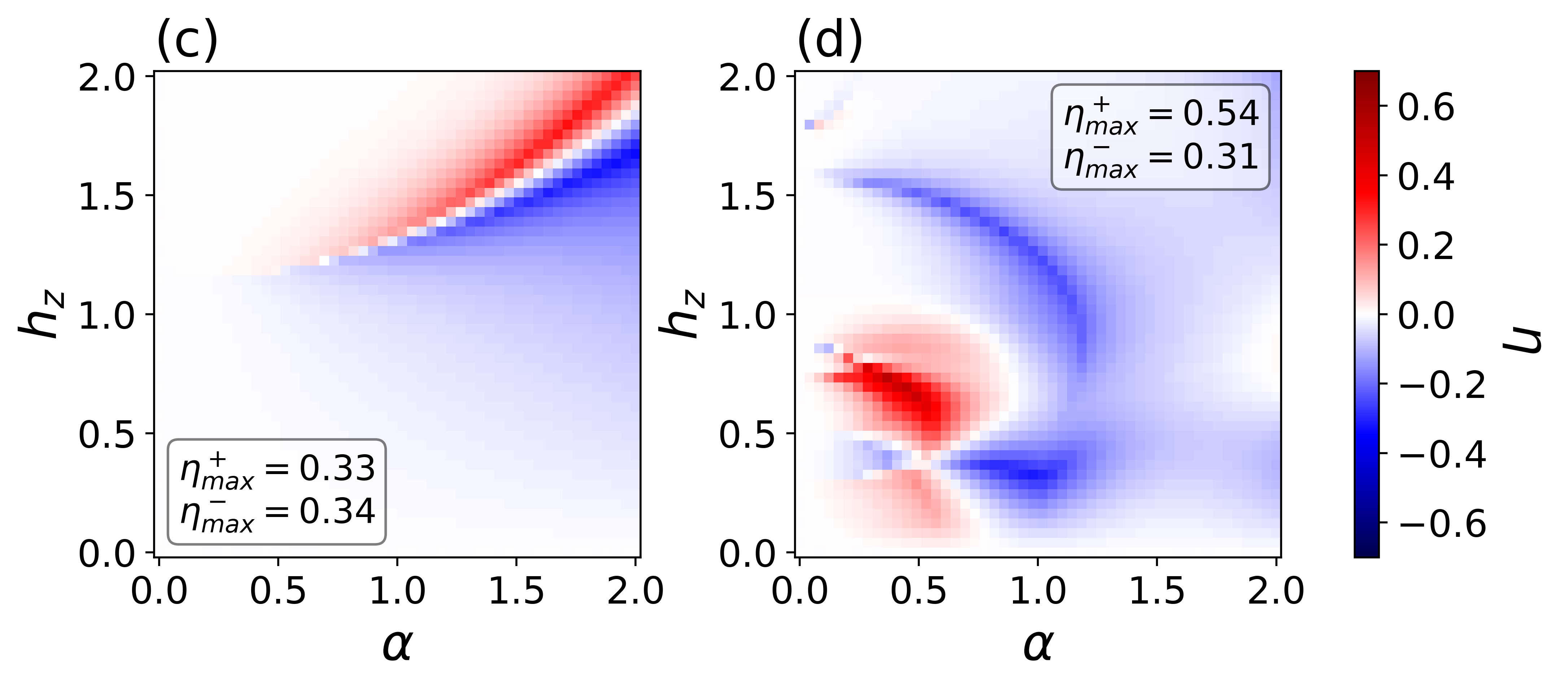}\label{}}
	\caption{The color map of efficiencies with a dense sampling of (a) $h_x$ and $h_z$ in the range $[0,2]$ for fixed $\alpha=0.9$ and (b) $\theta/\pi$ and $\alpha$ in the range $[0,2]$ for fixed magnitude of Zeeman field $h=0.9$. The color map of efficiencies with a dense sampling of $\alpha$ and $h_z$ for (c) $l=2L/100 $ and (d) $l=6L/100$. $k_x=0.5$ in all the calculations.
		The other parameters in the calculations are $L=150$, $a=1$, $t =3$, $\mu=1.5$ and $\Delta_0=0.3$.}
	\label{effLLNth}
\end{figure}

In our previous exploration, we focused on the system with the Zeeman field $h_z$ oriented perpendicularly to the surface of the JJs. Now, we extend our study to examine scenarios involving both out-of-plane and in-plane Zeeman fields. In Fig.\ref{effLLNth} (a), we select the SOC strength $\alpha=0.9$ and the transverse momentum $k_x=0.5$. By applying Zeeman fields with a fixed magnitude and varying directions in the $xz$-plane, we observe distinct strengths of the JDE. Notably, the pattern of efficiencies exhibits symmetry about the line $h_z=h_x$, as both $h_z$ and $h_x$ play equivalent roles in breaking the symmetry along the $y$-direction in equation \eqref{eq3}.
The peak magnitude of the efficiencies occurs at $(h_x,h_z)=(0.20, 0.33)$ with $\eta=-0.34$ or for the exchanged values of $h_x$ and $h_z$. Optimal values of $h_z$ and $h_x$ with other fixed parameters also exist. These findings suggest that by altering the direction of the magnetic field in addition to its magnitude, we can effectively adjust the efficiency of materials with $C_{1v}$.

To provide a clearer understanding of the impact of the direction of the Zeeman field on the efficiencies, we plotted the efficiencies as a function of angle $\theta$ and SOC in Fig.\ref{effLLNth} (b). Here, angle $\theta$ is defined by the angle-dependent Zeeman fields $h_x=h\cos \theta$ and $h_z=h\sin \theta$ with a fixed magnitude ($h=0.9$ in Fig.\ref{effLLNth} (b)). Interestingly, at angles $\theta=3\pi/4$ and $7\pi/4$, the efficiencies approach zero, which is independent of the SOC and the magnitude of the Zeeman fields (also shown in Fig.\ref{eff_pmhzhx} (c) and (d) with different magnitudes of Zeeman fields). This observation suggests that the CPRs exhibit equal magnitudes of critical currents in opposite directions at these angles, which can be explained by symmetry analysis. The Hamiltonian of the SOC and Zeeman parts can be expressed as
\begin{equation}
	\begin{aligned}
	H_{SZ}\left(k_x,k_y\right)=&\left(\alpha k_y+\frac{h}{\sqrt{2}}\right)\sigma_x+k_x \sigma_y \\
	&\qquad\qquad +\left(\alpha k_y-\frac{h}{\sqrt{2}}\right) \sigma_z,\\
	\end{aligned}
	\label{eq12}
\end{equation}
where $h_x=-h_z=h/\sqrt{2}$. Here we can define a symmetry operation $\mathcal{U}=\sigma_z R_y\left(\frac{\pi}{2}\right)$ where $R_{y}\left(\frac{\pi}{2}\right)=e^{-i\frac{\pi}{2}\frac{\sigma_{y}}{2}}$. The Hamiltonian equation \eqref{eq12} satisfies the relation $\mathcal{U}H_{SZ}\left(k_x,k_y\right) \mathcal{U}^{-1}=H_{SZ}\left(-k_x,-k_y\right)$. The accidental symmetry at the configuration of the Zeeman field prevents the appearance of JDE, illustrating the significance of symmetry breaking in finding systems with SDE.



Previously, discussions revolved around a fixed junction length of $l=5L/100$. Our current focus is on exploring how the junction length affects the JDE with a single transverse momentum of  $k_x=0.5$, and a Zeeman field oriented in the $z$-direction. In Fig.\ref{effLLNth} (c) and (d), we display the efficiencies for junction lengths of $l=2L/100$ and $l=6L/100$, respectively. As the junction length increases, there is a non-monotonic change in the maximum efficiency magnitude with the optimal values of the SOC and Zeeman field shifting (Fig.\ref{lengthex}). Longer junction lengths lead to more ``islands" with large efficiency magnitudes due to the increased number of ABSs moving into the gap. The joint contributions to the current from each ABS at a fixed $\varphi$ may change more dramatically than those with a small number of ABSs when other parameters change. In other words, the $0$-$\pi$-like transitions for more ABSs become more sensitive to changes in the other parameters, resulting in complicated distributions of positive and negative efficiencies. 

Comparing the results in Fig.\ref{effLLNth} (c), (d) and Fig.\ref{lengthex}, it is evident that the maximum efficiency magnitude initially rise and then decrease with an increase in junction length.
The peak efficiency magnitude is achieved at $l=4L/100$ with $\eta = 0.67$  for all the data we presented, as depicted in Fig.\ref{lengthex} (b).
This suggests that for a single $k_x$-channel, a greater number of ABSs may enhance the JDE for specific junction lengths, depending on the optimal SOC and Zeeman fields. However, the enhancement is not guaranteed to be monotonically increasing or even exist with an increasing number of ABSs. These findings imply that for a single $k_x$-channel, a comparatively shorter junction, featuring a particular joint contribution of ABSs, can offer a considerable efficiency at a higher critical current density than longer junctions. In other words, a higher proportion of the normal part with SOC and the Zeeman field does not necessarily to achieve the maximum magnitude of efficiency.
In conclusion, careful adjustment of the presented parameters is necessary to achieve the maximum magnitude of efficiency.  

\section{\label{sec4}Conclusions}

We conducted a numerical study on the JDE in planar JJs with $C_{1v}$ SOC and Zeeman fields. Our study built upon the experimental findings presented in \cite{chen2024edelstein}, which demonstrated the origins of JDE in planar \(T_{d}-MoTe_{2}\) JJs. Our results align with the analysis presented in \cite{chen2024edelstein}, highlighting the disappearance of JDE in relatively thick JJs.
Specifically, our investigations reveal an asymmetric distribution of ABS and a $0$-$\pi$-like transition with $\varphi_0$-shift in the quasi-2D JJs system with $C_{1v}$ SOC and an out-of-plane magnetic field. We explored the relationship between the $0$-$\pi$-like transition and the JDE, analyzed the single and joint effects of ABSs in a single transverse channel, and demonstrated an enhanced JDE for the joint effect of ABSs. By presenting the efficiencies as functions of various parameters, including strength of the SOC, Zeeman field magnitude and orientation, and junction length, we thoroughly investigated their influence on the efficiencies. In addition, we compared JJs with single and multiple transverse channels by examining their CPRs and efficiencies. Our findings indicate non-monotonic changes in efficiencies with respect to various parameters, consistent with the results presented in \cite{chen2024edelstein} when considering the efficiencies as a function of the magnetic field. We underscore the significance of symmetry breaking for JDE based on the results we presented. Our study showcases diverse possibilities for tuning the efficiencies using tunable SOC, Zeeman field, and length or width of the JJs. Based on our analysis, high efficiencies may be acquired by selected parameters or in a system with symmetric properties other than $C_{1v}$ symmetry.
These insights may provide valuable clues for discovering materials and designing Josephson diodes with a significantly large magnitude of efficiency.


\begin{acknowledgments}
We thank Hong-Tao He and Ying Liu for helpful discussions. This work was supported by the National Key R\&D Program of China (Grants No. 2022YFA1403700), NSFC (Grants No. 12141402, 12334002), the Science, Technology and Innovation Commission of Shenzhen Municipality (No. ZDSYS20190902092905285), Guangdong Provincial Quantum Science Strategic Initiative Grand No. SZZX2401001, the SUSTech-NUS Joint Research Program, and Center for Computational Science and Engineering at Southern University of Science and Technology.
\end{acknowledgments}

\nocite{*}

\bibliographystyle{apsrev4-1}
\bibliography{bibliography}

\appendix*

\section{Efficiencies with various parameters}

\begin{figure}[H]\centering
	\subfloat[]{\includegraphics[width= 0.5\linewidth,valign=b]{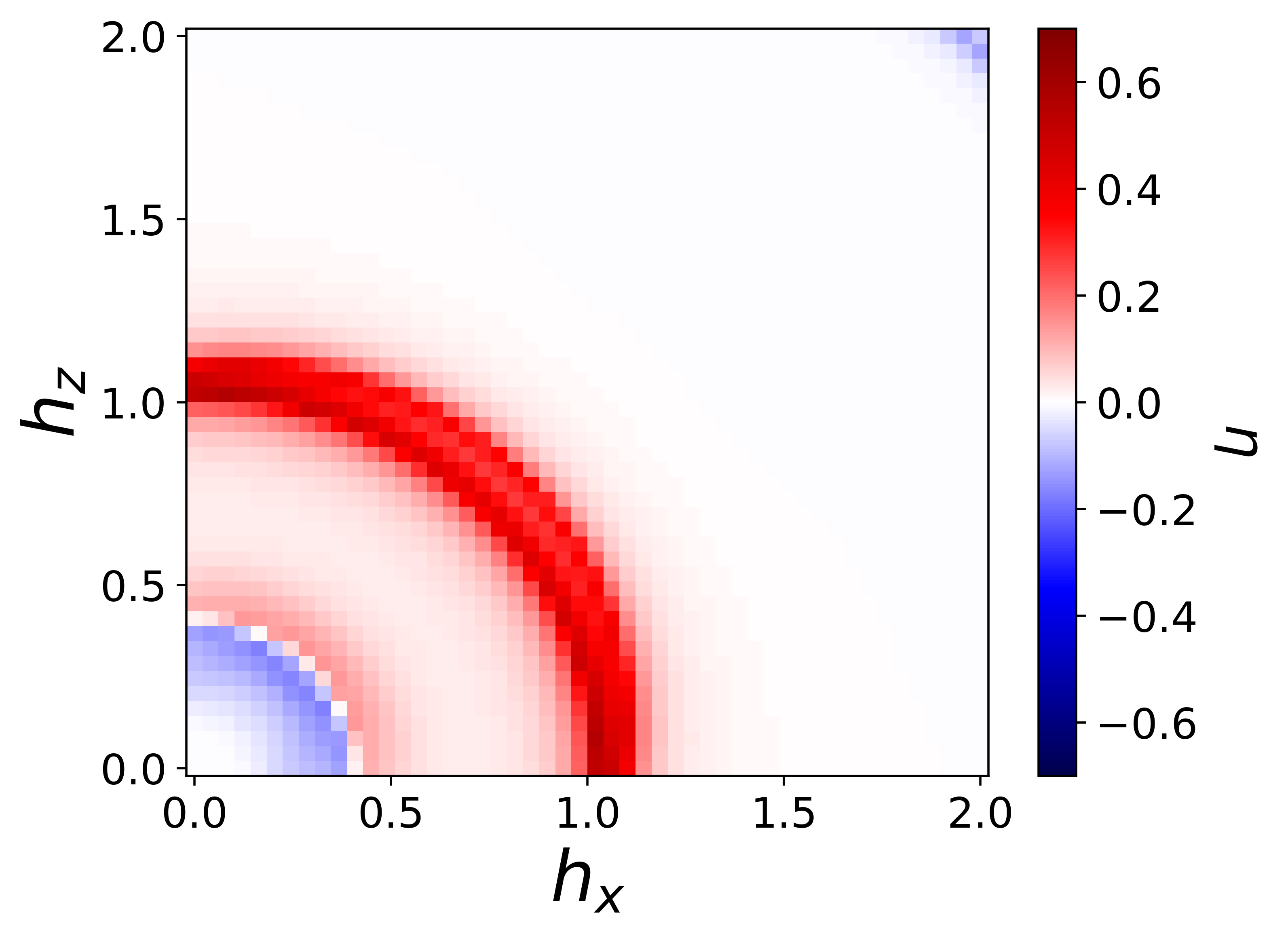}\label{}}
	\subfloat[]{\includegraphics[width= 0.5\linewidth,valign=b]{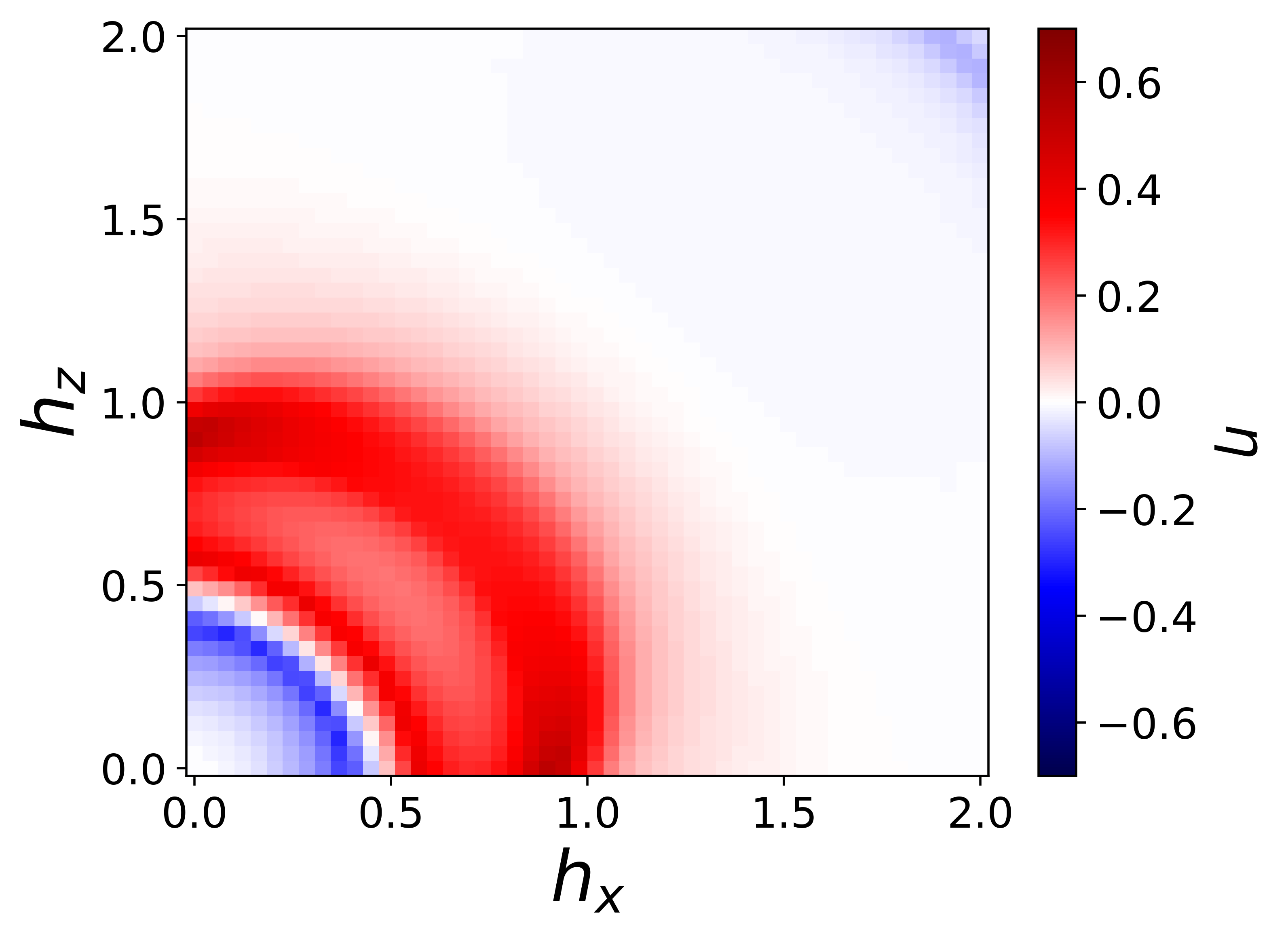}\label{}}\\
	\subfloat[]{\includegraphics[width= 0.5\linewidth,valign=b]{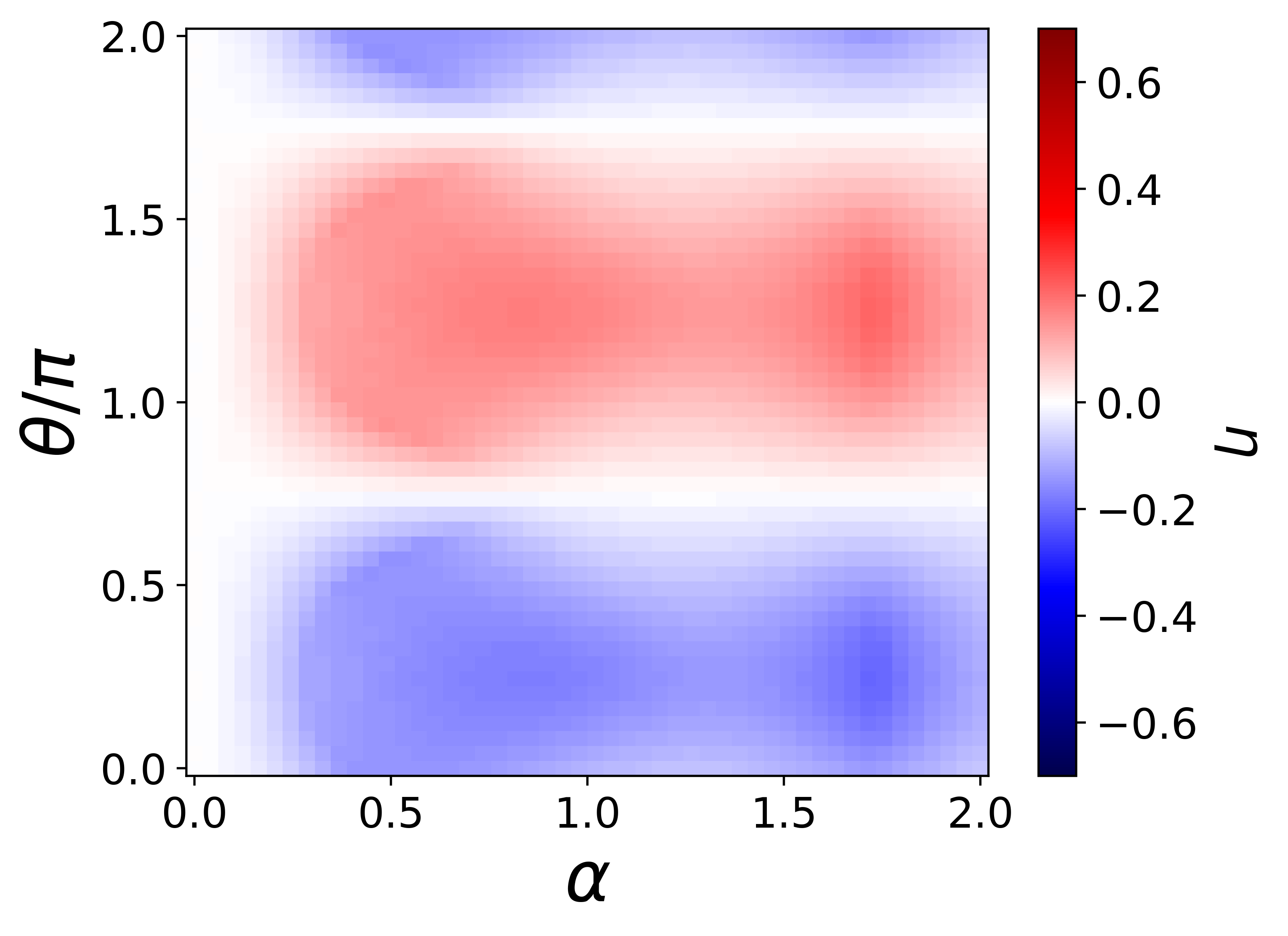}\label{}}
	\subfloat[]{\includegraphics[width= 0.5\linewidth,valign=b]{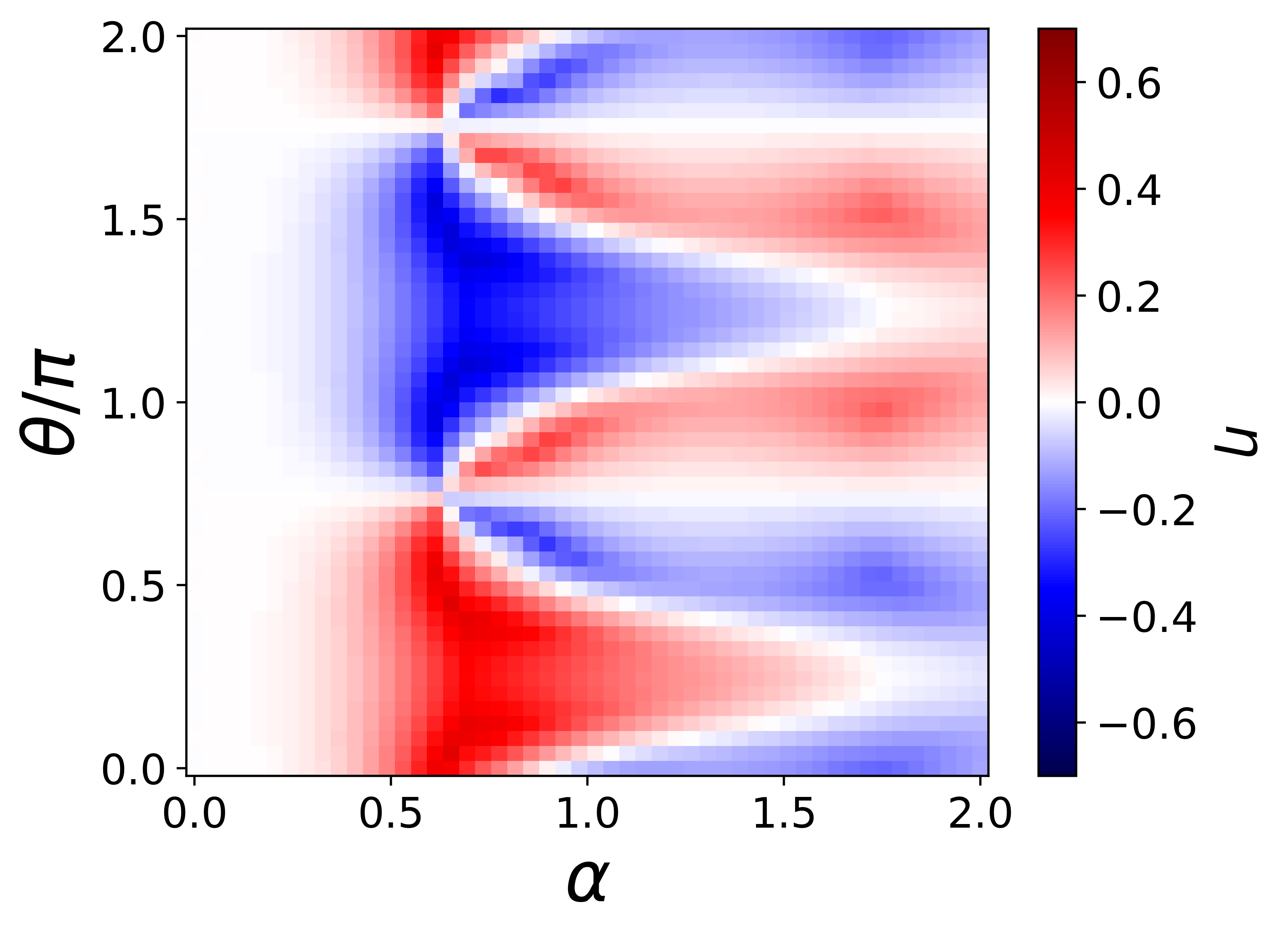}\label{}}
	\caption{The color map of efficiencies with a dense sampling of $h_x$ and $h_z$ in the range $[0,2]$ for (a) $\alpha=0.3$ and (b) $\alpha=0.6$. The color map of efficiencies with a dense sampling of $\theta/\pi$ and $\alpha$ in the range $[0,2]$ for fixed magnitude of Zeeman field (c) $h=0.3$ and (d) $h=0.6$. The other parameters in the calculation are $k_x=0.5$, $l=5L/100 $, $L=150$, $a=1$, $t =3$, $\mu=1.5$ and $\Delta_0=0.3$. The maximum magnitude of efficiencies occur at $h_z=0.08,h_x=1.02$ with $\eta=0.55$ (Fig.\ref{effLLNth} (a)), $h_z=0,h_x=0.90$ with $\eta=0.53$ (Fig.\ref{effLLNth} (a)) and $h_z=0.33,h_x=0.20$ with $\eta=-0.34$ (Fig.\ref{effLLNth} (b)) or for exchanged $h_z$ and $h_x$.}
	\label{eff_pmhzhx}
\end{figure}

\begin{figure}[H]\centering
	\subfloat[]{\includegraphics[width= 0.50\linewidth,valign=b]{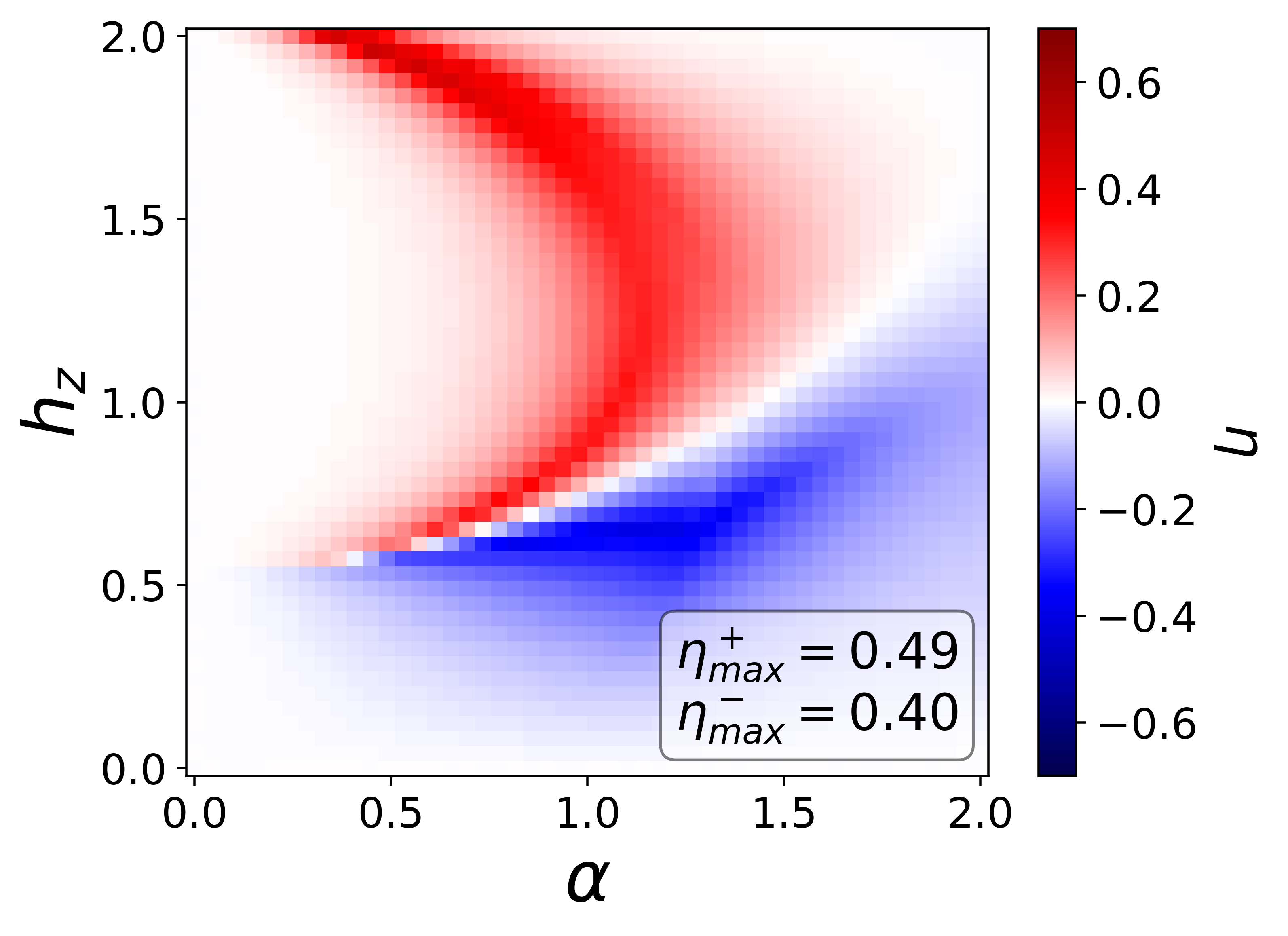}\label{}}
	\subfloat[]{\includegraphics[width= 0.50\linewidth,valign=b]{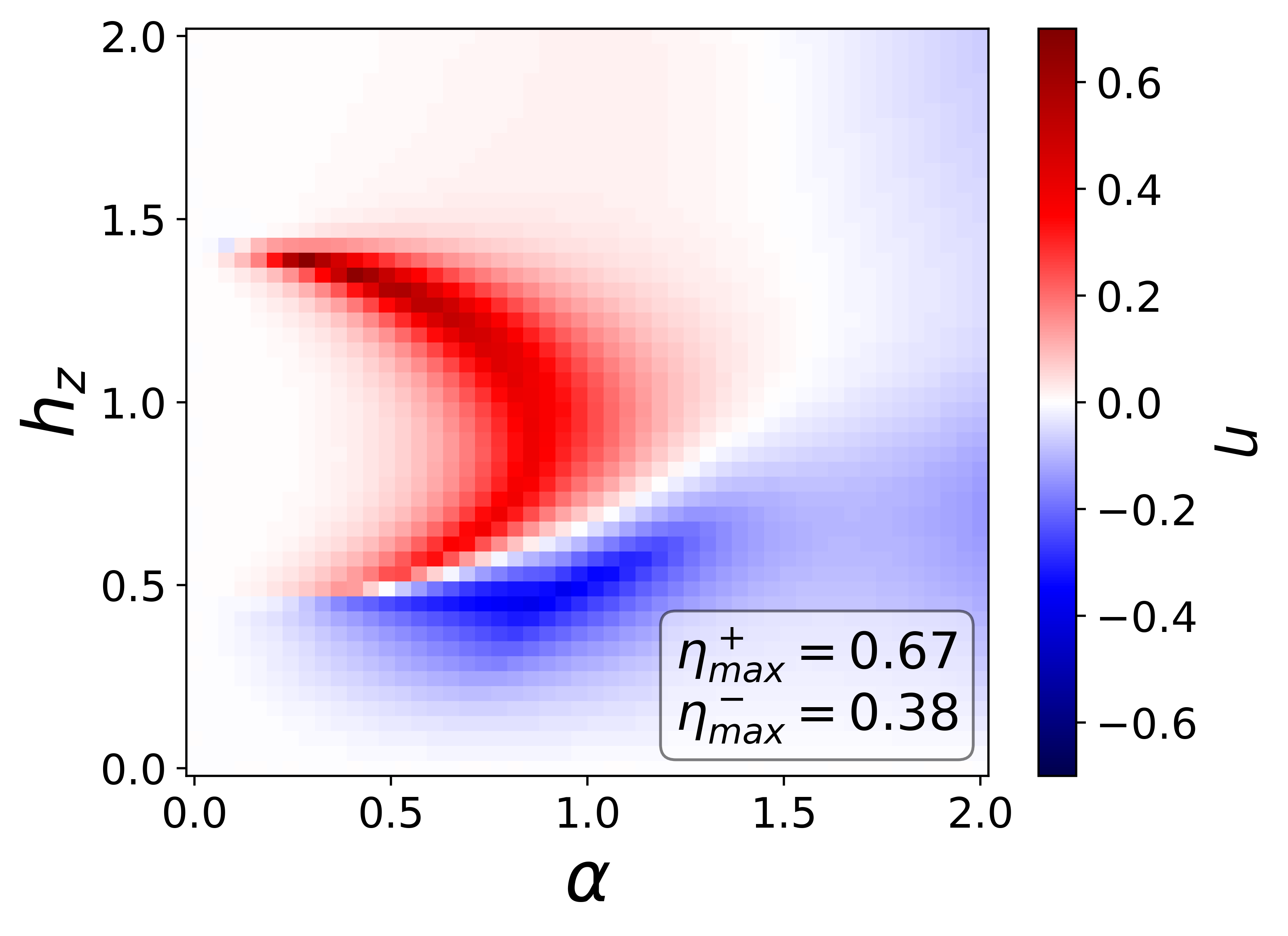}\label{}}\\
	\subfloat[]{\includegraphics[width= 0.50\linewidth,valign=b]{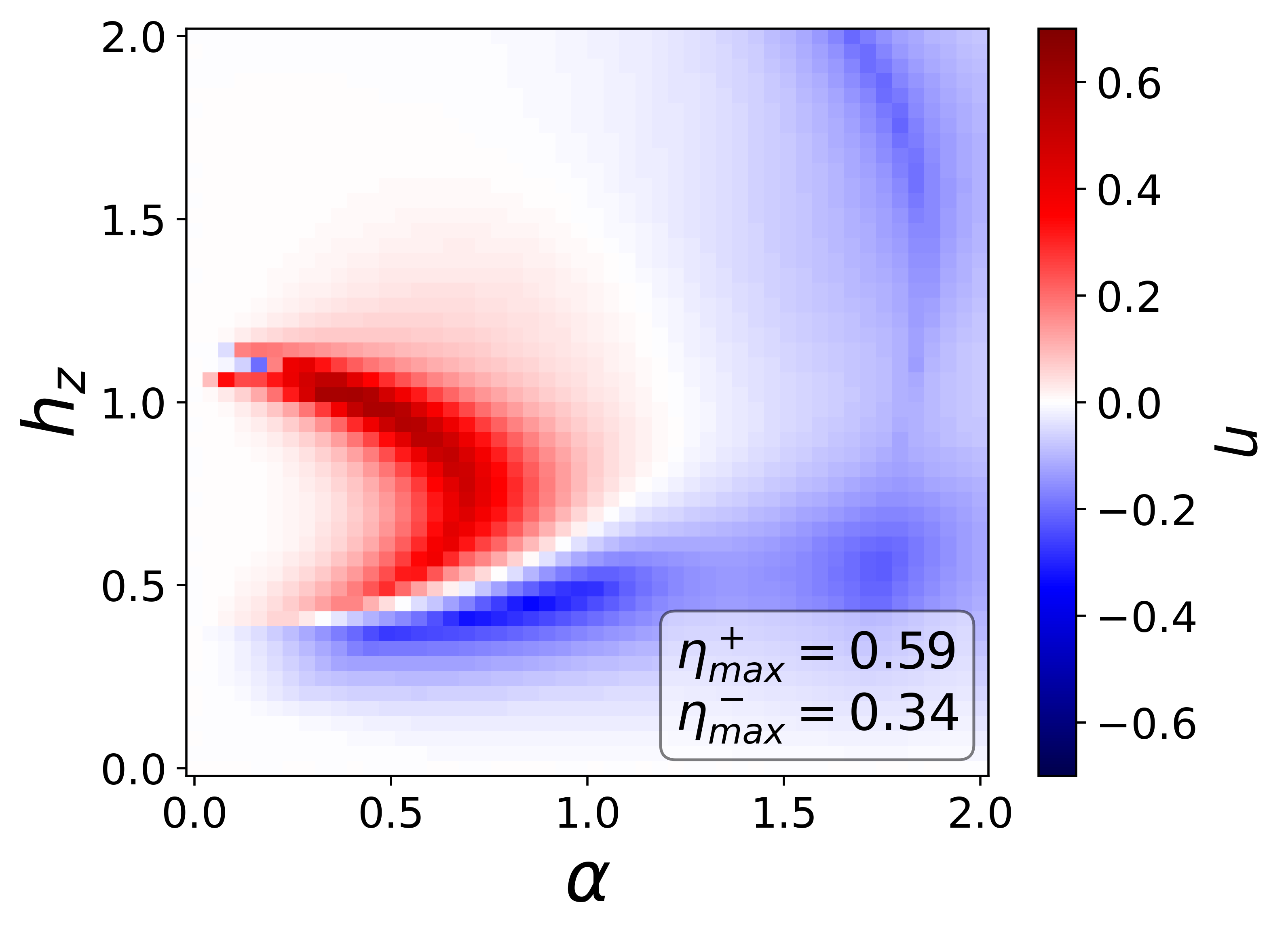}\label{}}
	\subfloat[]{\includegraphics[width= 0.50\linewidth,valign=b]{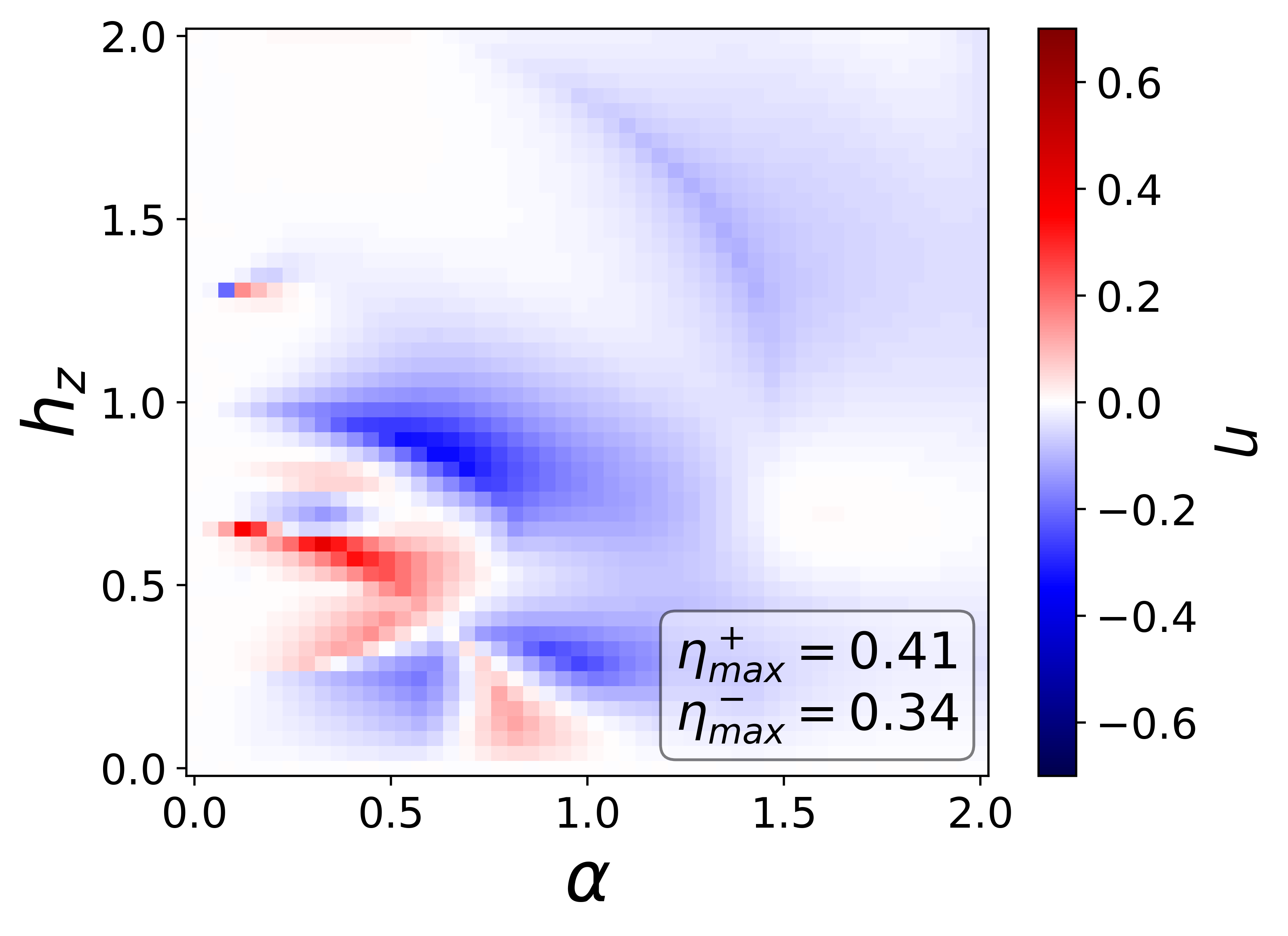}\label{}}\\
	\subfloat[]{\includegraphics[width= 0.50\linewidth,valign=b]{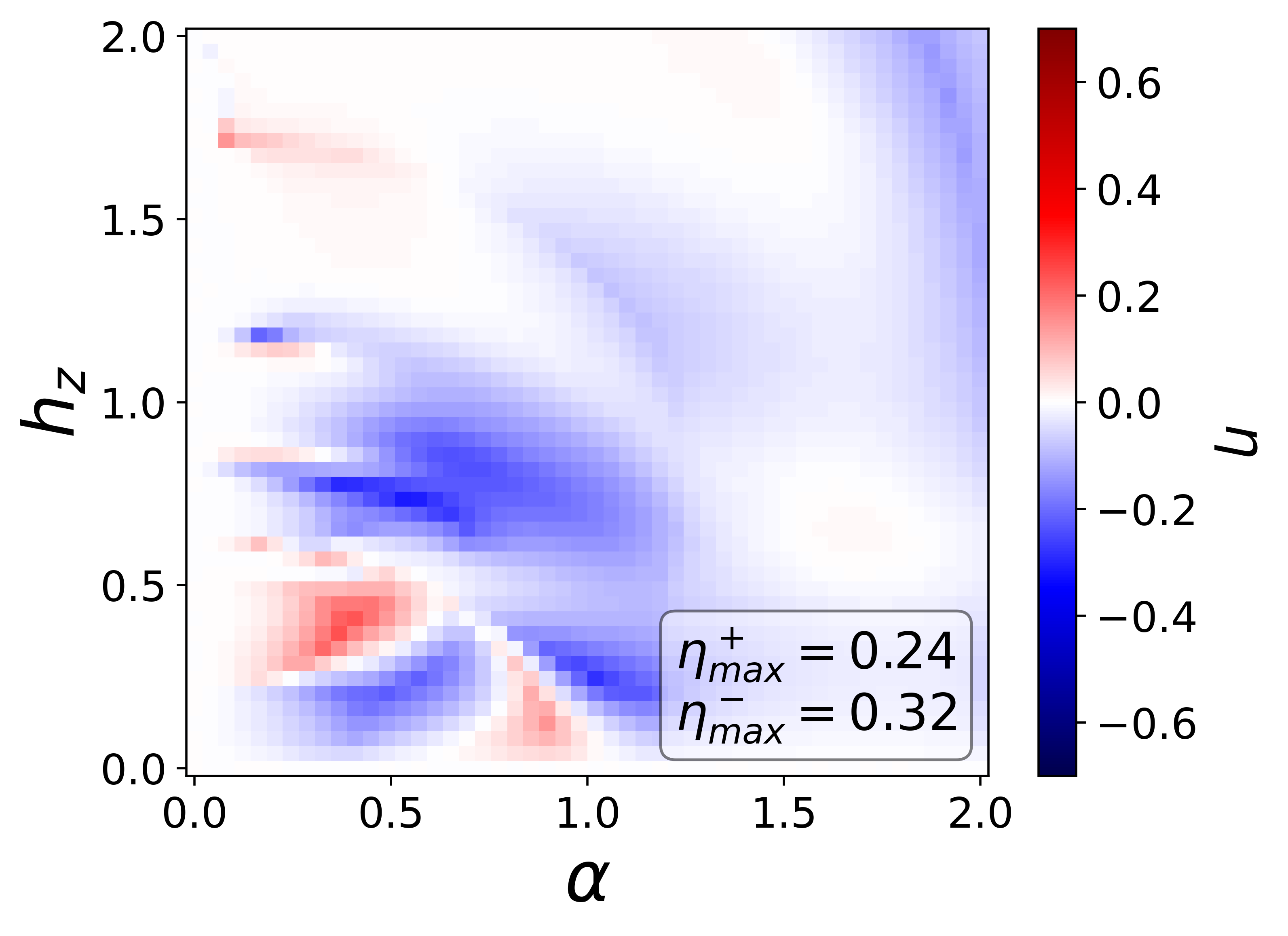}\label{}}
	\subfloat[]{\includegraphics[width= 0.50\linewidth,valign=b]{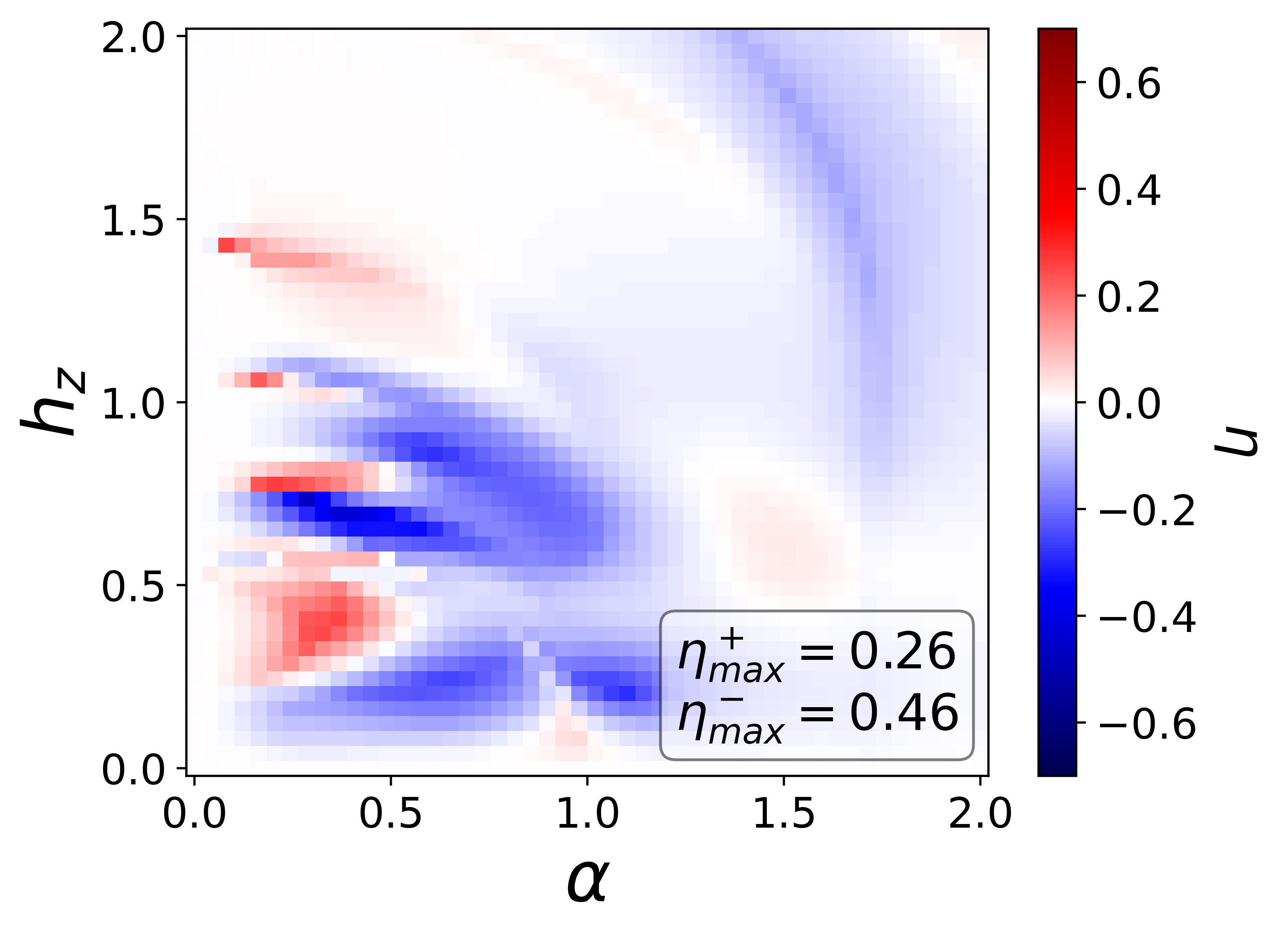}\label{}}
	\caption{ The color map of efficiencies with a dense sampling of $\alpha$ and $h_z$ for (a) $l=3L/100 $, (b)$l=4L/100 $, (c) $l=5L/100 $, (d) $l=7L/100 $, (e) $l=8L/100 $ and (f) $l=9L/100$. $k_x=0.5$ in all the calculations. The other parameters in the calculation are $k_x=0.5$, $L=150$, $a=1$, $t =3$, $\mu=1.5$ and $\Delta_0=0.3$.}
	\label{lengthex}
\end{figure}

Here, we present additional color maps of efficiencies as functions of different parameters to support the discussion in Sec. \ref{D}. In Fig.\ref{eff_pmhzhx} (a) and (b), the efficiency is depicted as a function of $h_x$ and $h_z$ with varying strengths of SOC at $\alpha=0.3$ and $0.6$. In Fig.\ref{eff_pmhzhx} (c) and (d), the efficiency is plotted as a function of $\theta$ and $\alpha$ for different Zeeman field strengths at $h=0.3$ and $0.6$. The same symmetric properties are shown, as discussed in Sec. \ref{D}. In Fig.\ref{lengthex}, additional results of the efficiency as a function of the SOC and Zeeman fields with different junction lengths are provided.


\end{document}